\documentstyle[12pt]{article}
\newcommand{\be}{\begin{equation}}
\newcommand{\ee}{\end{equation}}
\newcommand{\bea}{\begin{eqnarray}}
\newcommand{\eea}{\end{eqnarray}}

\def\E{\> = \>}

\begin{document}
\thispagestyle{empty}
\setcounter{page}{0}

\vspace{1cm}

\begin{center}
{\large\bf Proton Polarization Shifts in Electronic and Muonic Hydrogen}

\vspace{1cm}

\renewcommand{\thefootnote}{\fnsymbol{footnote}}

R.~Rosenfelder \footnote[1]{E-mail address: rosenfelder@psi.ch}

\renewcommand{\thefootnote}{\alph{footnote}}

\vspace{1cm}

Paul Scherrer Institute, CH-5232 Villigen PSI, Switzerland

\end{center}

\vspace{5cm}
\begin{abstract}
\noindent
The contribution of virtual excitations to the energy levels of electronic
and muonic hydrogen is investigated combining a model-independent
approach for the main part with quark model predictions for the remaining
corrections. Precise values for the polarization shifts are obtained in the 
long-wavelength dipole approximation by numerically integrating over measured 
total photoabsorption cross sections. These unretarded results
are considerably reduced by including retardation effects 
in an approximate way since the average momentum transfer (together with 
the mean excitation energy) turns out to be larger than usually assumed.
Transverse and seagull contributions are estimated in a simple
harmonic oscillator quark model and found to be non-negligible.
Possible uncertainties and improvements of the final results are discussed.

\end{abstract}

\vspace{0.5cm}
\hspace{0.2cm} PACS numbers: 12.20.Ds, 12.39.Jh, 14.20.Dh, 36.10.Dr

\newpage

\setcounter{equation}{0}

\noindent
{\bf 1.} There has been tremendous progress in the laser spectroscopy
of hydrogen and deuterium atoms \cite{Exp} which now are even sensitive
to small nuclear and proton structure effects. One of these - traditionally 
the least understood - is the virtual excitation of the nucleus which acts 
back on the
bound lepton. While the effect in deuterium is comparatively large and has
been evaluated theoretically with increasing sophistication and reliability
\cite{deutpol}, the proton polarization shifts have not received much attention
up to now. Khriplovich and Sen'kov \cite{KhSe} have estimated the shift
in the electronic $1S$-state as $ -71 \pm 11 \pm 7 $ Hz using values for the 
static proton polarizabilities and assuming a mean excitation energy of 
$\sim 300$ MeV. They attribute the quoted errors to the use of a 
relativistic approximation for the electron and to the experimental values of
the polarizabilities. However, experience gained in many decades of
nuclear polarization calculations has told us that the use of an average 
excitation energy (the ``closure approximation'') can be a considerable source of 
uncertainty unless one calculates it precisely. In addition, one has to make
sure that other ingredients to the polarization shift (higher multipoles,
transverse excitations) are well under control before a definite answer
can be given.
It is the purpose of this note to re-evaluate the polarization 
shift without the questionable use of a mean excitation energy
and other simplifying assumptions. 
Since an experiment is in progress at PSI 
to measure the Lamb shift in muonic hydrogen \cite{PSI-R-98-03}
we will also evaluate the polarization shift for this case. Actually, with the
experimental accuracies achievable in the near future, it turns out that
the proton polarization shifts are
of much greater importance for muonic than for electronic hydrogen.

\vspace{0.3cm}
\noindent
{\bf 2.}
For light electronic and muonic atoms the
energy shift due to virtual excitations can be written as 
an integral over the forward virtual Compton amplitude. This quantity in turn
may be expressed by its imaginary part, i.e. the  structure functions
$W_{1/2}(\nu,Q^2)$ which are measurable in the inclusive reaction
$ e + p \longrightarrow e' + X $. In the absence of detailed experimental
information for all relevant values of momentum transfer $Q$ 
and energy transfer $\nu$ it is customary to apply the long wavelength
(or unretarded dipole) approximation which should be valid for
$ \bar Q \left < r^2 \right >^{1/2}  \ll 1 $ where 
$ \left < r^2 \right >^{1/2} $ is the root-mean-square radius of the proton 
and $\bar Q$ an average
momentum transfer. In this limit it is possible to express the structure 
functions by the experimentally measured photoabsorption cross section 
$\sigma_{\gamma}(\nu)$. Bernab\'eu and Ericson have derived in this way the 
following expression for the energy shift \cite{BeEr} 
\be
\Delta E_{nl} \E - \frac{\alpha}{2 \pi^2} \, \frac{| \psi_{nl}(0)|^2}{m}
\, \int_0^{\infty} d \nu \> \sigma_{\gamma}(\nu) \, f\left ( \frac{\nu}{2 m}
\right )
\label{Delta E unret}
\ee
with
\be
f\left ( \frac{\nu}{2 m} \right ) \E \frac{8 m^2}{\pi} \int_0^{\infty} d Q^2
\> \frac{1}{Q^2} \, \int_0^Q d \xi \> \sqrt{Q^2 - \xi^2} \, 
\frac{Q^2 + 2 \xi^4/Q^2}{(\nu^2 + \xi^2) (Q^4 + 4 m^2 \xi^2)} \> .
\label{BE function}
\ee
\noindent
Here $\alpha$ is the fine-structure constant, $m$ the lepton mass and
$ \psi_{nl}(0)$  the lepton wave function at the origin which is only non-zero
for $S$-states. The correction factor accounting for the variation of the 
wave function
over the proton radius can be safely neglected at this level of accuracy.
We first note that it is possible to perform all integrations
in eq. (\ref{BE function}) \footnote{This has also been noted before 
\cite{Ber} .} and to obtain
\be
f(x) \E - 2 - 2 \ln (4 x) + \frac{1 + 2 x^2}{x^2} \, \left [ \, 
\sqrt{x^2 + x} \ln \frac{\sqrt{x^2 + x} + x}{\sqrt{x^2 + x} - x} + g(x) \, 
\right ]
\ee
with 
\be
g(x) \E \sqrt{|x^2 - x|} \, \left [ \> \Theta(1-x) \>
2 \arctan \frac{\sqrt{x - x^2}}{x} \, - \, \Theta(x-1) \> 
\ln \frac{x +\sqrt{x^2 - x}}{x - \sqrt{x^2 - x}} \> \right ] \> .
\ee
The function $f(x)$ has quite different limits for small and large arguments:
\bea
f(x) &\longrightarrow & \pi x^{-3/2} \hspace{4.3cm} {\rm for} \> x \to 0 
\label{nonrel}\\
&\longrightarrow & \frac{5}{4x^2} \left [ \,  \ln (4x) + \frac{19}{30} \, 
\right ]  \hspace{2cm} {\rm for} \> x \to \infty           
\label{rel}
\eea
which makes a crucial difference between 
nuclear polarization shifts, say, in muonic deuterium and in electronic 
hydrogen.
Krhriplovich and Sen'kov have only retained the leading logarithm in 
eq. (\ref{rel}) which is justified considering the much greater error
which comes from pulling out the logarithmic term evaluated at the excitation
energy of the $\Delta(1232)$ and expressing the remaining integral in terms of 
the sum of polarizabilities
\be
\bar \alpha + \bar \beta \E \frac{1}{2 \pi^2} \int_0^{\infty} d\nu \> 
\frac{\sigma_{\gamma}(\nu)}{\nu^2} \> .
\label{polarizabilities}
\ee

\vspace{0.3cm}
\noindent
{\bf 3.} In view of the well-known deficiencies of the closure approximation
we have decided to evaluate the polarization shift by integrating numerically
over the experimentally measured photon absorption cross section.
We have taken the recent Mainz data \cite{Mainz} from $\nu = 200 - 800$ MeV,
the older Daresbury data \cite{Daresbury} from $\nu = 800 - 4215$ MeV 
and the parametrization 
$ \sigma_{\gamma}(\nu) = 96.6 \, \mu {\rm b} + 70.2 \, \mu  {\rm b} \> 
{\rm GeV}^{1/2} /\sqrt{\nu} \> $ above $4215$ MeV. Below $200$ MeV four 
angular distributions in $ \gamma p \to \pi^+ n$ 
(from Fig. 1 in ref. \cite{Han})
have been converted to total cross sections and
the threshold behaviour has been parametrized as 
$ \sigma_{\gamma}(\nu) = 18.3 \, \mu  {\rm b} \> {\rm MeV}^{-1/2} \,  
\sqrt{ \nu - \nu_{\rm th}} \> $.
The numerical integration was done by simple Simpson integration over 
linearly interpolated data points and 
the whole procedure was checked
by evaluating eq. (\ref{polarizabilities}). We obtained
\be 
\bar \alpha + \bar \beta \E 13.75 \cdot 10^{-4} \> \> {\rm fm}^3
\label{num polarizabilties}
\ee
in good agreement with a recent analysis \cite{BGM}  but smaller than the value
$ 14.2 \cdot 10^{-4} \> \> {\rm fm}^3 $
used in Ref. \cite{KhSe}. This is due to the fact that the older
cross sections which have been analysed in Ref. \cite{DaGi} 
are systematically higher than the new Mainz data.

\noindent
The {\it unretarded} results for the electronic (e) and muonic ($\mu$) 
polarization shifts are then
\bea
\Delta E_{nS}^e &=& - \frac{106.5}{n^3} \> \> {\rm Hz} 
\label{unretarded e shift}\\
\Delta E_{nS}^{\mu} &=& - \frac{202}{n^3} \> \> \mu{\rm eV} \> .
\label{unretarded mu shift}
\eea
In view of the small electron mass 
it is not surprising that the relativistic approximation
(\ref{rel}) (including the constant term) agrees with the exact result
to more than 4 digits.   
Equation (\ref{unretarded e shift}) is larger than the estimate of ref. 
\cite{KhSe} because the integral over virtual excitations gets contributions
well above the $\Delta$-resonance. This can be easily seen from any graph of
the total photoabsorption cross section versus photon energy but made more
quantitative by asking for the value of the mean excitation energy $\bar \nu $
which
-- when substituted into the logarithm of eq. (\ref{rel})~-- gives the same 
result for the shift. We find $\bar \nu \simeq 410 $ MeV 
both for electronic and muonic hydrogen.
Therefore predictions assuming that only the
$\Delta$ isobar contributes to virtual excitations of the proton \cite{FMS} 
are not very reliable. In the muonic case the approximation
(\ref{rel}) still gives more than $96$ \% of the exact numerical result whereas
the non-relativistic approximation (\ref{nonrel}) overestimates it by
more than $40$ \%. This is, of course, consistent with the fact that 
the mean excitation energy is nearly four times larger than the muon mass.

\vspace{0.3cm}
\noindent
{\bf 4.} The large value of the mean excitation energy also casts some doubt
on the use of the unretarded dipole approximation. For example, in a simple
constituent harmonic oscillator quark model one naively expects
an associated mean momentum transfer of 
$ \bar Q \simeq \sqrt{  M_{\rm quark} \bar \nu} \simeq 350$ MeV. 
This will have an appreciable effect when inserted in the 
elastic form factor which, e.~g. in the dipole approximation is
given by $ F_0 (Q^2) \simeq 1/(1 + Q^2/Q_0^2)^2 $ with
$ Q_0^2 = 0.71 \>  {\rm GeV}^2 $. Inelastic transition form factors to 
low-lying resonances have approximately the same $Q$-dependence apart from
threshold factors which are characteristic for the specific angular momentum
of the resonance (see below for a non-relativistic example).
Therefore one obtains a rough estimate for the effect of
retardation when the square of the elastic form factor $ F_0^2 (Q^2)$ 
is inserted in the $Q^2$-integral of eq. (\ref{BE function}). Instead of
the dipole form factor we
have employed a more realistic parametrization of the charge form factor
of the proton given in eq. (9) and table 3 of ref. \cite{Simon}.
A careful numerical evaluation of the remaining double integral 
then gives for the {\it retarded} polarization shifts 
\bea
\Delta E_{nS}^e &=&- \frac{88.9}{n^3} \> \> {\rm Hz} 
\label{retarded e shift} \\
\Delta E_{nS}^{\mu} &=& - \frac{112}{n^3} \> \> \mu{\rm eV} \> .
\label{retarded mu shift}
\eea
Use of the dipole form factor changes the numerical values to $89.3$
and $114$ for electron and muon, respectively.
Equation (\ref{retarded e shift}) is nearly $20 \%$ smaller in magnitude 
than the unretarded result and the polarization shift in muonic hydrogen 
gets almost halved due to retardation effects. This 
reduction can be translated into a mean momentum transfer which -- 
when inserted into the square of the elastic form factor -- cuts the
unretarded values by just this amount. In this way one obtains
$ \bar Q \simeq 180$ and $ 340 $ MeV/c for the electronic and muonic case 
respectively. 

\vspace{0.3cm}
\noindent
{\bf 5.}
At present this seems to be the best model-independent estimate
for the polarization shifts in hydrogen but it still neglects higher
multipoles and relies on a standard but not very well tested procedure 
to correct the unretarded dipole approximation.
Contrary to the deuteron case
where excellent potential models are available, a reliable 
(relativistic~) model of the nucleon does not exist 
to calculate all these contributions. Here we take the simple non-relativistic
harmonic oscillator quark model to estimate them. In this model it is easy to 
evaluate analytically the longitudinal (inelastic) structure function
\be
S_L(\nu,q) \E \sum_{N=1}^{\infty} \, \delta \left ( \nu - N \omega - 
\frac{q^2}{6 M_{\rm quark} }\right ) \, \frac{y^N}{N !} \, e^{-y}
\label{long}
\ee
as well as the transverse one \footnote{The structure functions $S_{L/T}$ are 
linear combinations of the usual $W_{1/2}$
and are more convenient in the framework of ref. \cite{Ros} which
uses the energy transfer $\nu$ and 
the three-momentum transfer $ q \equiv | {\bf q}| $ as variables whereas
ref. \cite{BeEr} employs the invariants  $\nu$ and $Q^2$.}
\be
S_T(\nu,q) \E \frac{q^2}{2 M_{\rm quark}^2} \, S_L(\nu, q) + 
\frac{2 \omega}{3 M_{\rm quark}} \sum_{N=1}^{\infty} \, \delta \left ( 
\nu - N \omega - \frac{q^2}{6 M_{\rm quark} }\right ) \, 
\frac{y^{N-1}}{(N-1)! } \, e^{-y} \> .
\label{trans}
\ee
Here $ y = (q  b)^2/3 \> $ where $ b$ is the oscillator length and 
$ \omega = 1/(M_{\rm quark} b^2) $ the harmonic oscillator frequency.
In this model the excitation spectrum consists of sharp lines at 
$ N \omega$ shifted by the recoil energy of the proton
and the elastic form factor is gaussian ($ F_0^2(q) = \exp(-y) $)
so that the proton rms radius is directly given by the
oscillator length $b$. It should be noticed that the first and the second 
term in eq. (\ref{trans}) come from  excitations by the spin and 
the convection current, respectively. 
Note also that the above structure functions fulfill Siegert's theorem
$ \lim_{q \to 0} S_L(\nu,q)/q^2 = \lim_{q \to 0} S_T(\nu,q)/(2 \nu^2) = 
\sigma_{\gamma}(\nu)/(4 \pi^2 \alpha \nu) $ and that all excitations 
are indeed multiplied by the square of the elastic form factor.
In the low-$q$ limit the 
first excited state exhausts the dipole absorption cross section
which is not a very realistic feature of the model. Other shortcomings are
the well-known inability \cite{Holstein} of the harmonic oscillator quark
model to reproduce both the empirical rms-radius 
($ \> \left < r^2 \right >^{1/2} = 0.86 $~fm ) 
and the polarizabilities (\ref{polarizabilities}) if the constituent
quark mass is fixed to  $M_{\rm quark} = M_{\rm proton}/3$. 
We also make this choice because the masses of the constituents should add 
up to the total mass in a consistent non-relativistic treatment. In addition,
this value gives the correct recoil energy and leads 
to a reasonable result for the magnetic moment of the proton. 
This leaves only the harmonic oscillator length as free parameter
which we have fixed in such a way that
the results from the unretarded dipole approximation are obtained. In this 
way the quark model results can be directly compared with the shifts 
evaluated in the model-independent approach.
It is quite obvious that a {\it non-relativistic} model becomes inadequate 
for excitation energies and momenta of the order of the constituent mass.
However, here our aim only is to obtain a rough estimate for the remaining 
corrections beyond the retarded dipole approximation and for this purpose
the harmonic oscillator quark model may be not totally useless.

To obtain a quantitative estimate we have inserted
the analytic expressions (\ref{long}, \ref{trans})
into the formulae in the Appendix of ref.
\cite{Ros} (which keep the relativistic kinematics for the lepton), 
summed over oscillator shells up to $ N_{\rm max} = 20 $ 
and integrated numerically over the momentum transfer up to 
$q_{\rm max} = 3000$ MeV. Special care has to be exercised because of 
the apparent singularity in the transverse weight function at $ q = 0 $ 
which is canceled by the seagull contribution. The latter one is required 
for a gauge-invariant treatment of non-relativistic systems. Other 
numerical difficulties arise 
from the very different scales of electron mass  and mean
excitation energy  which requires very high 
numerical accuracy (up to $ 60 \times 72$ gaussian points compared to 
$ 6 \times 72$ in the muonic case) and from
the slow convergence ($ \sim 1/N_{\rm max}^2$) of the sum over excitations 
for the spin current which peaks at high momentum, i.e. high $N$ . The latter 
problem has been overcome by an analytic resummation and the 
overall numerical stability has been checked by varying 
$q_{\rm max}$, $N_{\rm max}$ and the number of 
gaussian integration points.
The results of the calculation are collected in Table 1. 
As expected the  $b$-values are too small to account for the proton radius
while the $\omega$-values ($ \sim 280 $ MeV) 
seem reasonable for low-lying nucleonic excitations. A simple way to cure these
deficiencies is to attribute the gaussian form factor to the quark core and
to introduce an extra formfactor $ 1/(1 + \beta^2 q^2) $ for the meson cloud
which surrounds the core and brings the rms-radius of the model in agreement
with the experimental value \cite{BHY}. Although this is rather {\it ad hoc} 
and theoretically not very appealing we have included this variant 
also in Table 1. 
Note that the unretarded dipole approximation (and therefore the
value obtained in ref. \cite{KhSe}) also includes some
transverse excitations : eq. (\ref{rel}) would read $ (\ln 4x + 1)/x^2 $ 
if only longitudinal excitations are kept. 
While the longitudinal excitations (including all multipolarities) 
are seen to dominate, virtual transverse excitations induced by 
the currents cannot be neglected since the mean momentum transfer is 
of the order of the constituent quark mass. 
This is particularly important
for the spin current in muonic hydrogen because its contribution grows 
with momentum transfer. 

\vspace{0.3cm}

\begin{center}
\begin{tabular}{|l|c|c|}  \hline
                      &                      &                     \\
                      & electronic hydrogen ~~[Hz]~~  
                      & muonic hydrogen ~~[$\mu$eV]~~ \\
                      & ~~(A)~~~~~~~~~(B)     & ~~(A)~~~~~~~~~(B)     \\
                      &                      &                     \\ \hline
                      &                      &                     \\
 unretarded dipole (input)    & ~~~~- 106.4 & ~~~- 201.8    \\
 retarded dipole      & ~- 94.3~~~~~- 92.1   & - 131.6~~~~~- 121.1  \\ 
                      &                      &                     \\ \hline
                      &                      &                      \\
 full longitudinal    & ~~- 78.2~~~~~~- 75.9   & - 120.2~~~~~- 108.8   \\ 
 spin current         & ~~~- 7.2~~~~~~~- 3.8  & ~ - 39.1~~~~~~~- 20.9      \\
 convection current + seagull & ~~- 19.2~~~~~- 18.9  & ~- 16.9~~~~~~- 15.4 \\
 total                & - 104.6~~~~~- 98.6   & - 176.2~~~~~- 145.1     \\
                      &                      &                  \\ \hline \hline
                      &                      &                       \\ 
 correction to retarded dipole & ~- 10.3~~~~~~- 6.5   & - 44.6~~~~~~- 24.0 \\
                      &                      &                       \\ \hline
\end{tabular}

\end{center}

\vspace{0.1cm}
\noindent
{\bf Table 1} : Polarization shifts to the $1S$ level in electronic and muonic
hydrogen evaluated in the harmonic oscillator quark model. The 
parameters ($ M_{\rm quark} = 312.8 $ MeV, 
$b = 0.657 $ fm for electronic hydrogen, $b = 0.674 $ fm for muonic hydrogen) 
have been fitted to give the unretarded dipole approximation
and the corresponding results are given under the heading (A).
In case (B) an additional meson-cloud form factor has been introduced to 
reproduce the experimental proton radius.

\vspace{0.3cm}
\noindent
The ad-hoc introduction of the meson-cloud form factor 
reduces all contributions and brings the retarded dipole 
approximation more in accord with the calculation employing realistic form 
factors.

\vspace{0.5cm}

\noindent
{\bf 6.} In conclusion, we have evaluated the proton polarization shifts in 
electronic and muonic hydrogen in a fairly model-independent way by
integrating over the experimental photoabsorption cross section and 
accounting for retardation by use of the empirical elastic form factor.
The remaining contributions (mostly from 
transverse excitations) have been estimated in a simple harmonic
oscillator quark model and are therefore rather model-dependent 
and uncertain. Since it is physics in the resonance region which
dominates these contributions, the theoretical situation will probably 
remain so unless better experimental information from
inclusive $e p$ - scattering in this region is available.
For our final values we add the corrections (B) listed in Table 1 to 
eqs. (\ref{retarded e shift}, \ref{retarded mu shift}) and assign an 
error to them which covers the values obtained in case (A). 
This seems reasonable and prudent in view of the mentioned uncertainties and 
the inadequacy of the harmonic oscillator quark model.
In addition, we take the difference between the retarded dipole result
obtained with realistic form factors and the one with the gaussian form factor
for the core and monopole form factor for the meson cloud
as error estimate for the model-independent contribution.
Adding the errors linearly we obtain in this way our final result 
for the polarization shifts
\bea
\Delta E_{nS}^e &=& - \frac{95 \pm 7}{n^3} \> \> {\rm Hz} 
\label{final e shift} \\
\Delta E_{nS}^{\mu} &=& - \frac{136 \pm 30}{n^3} \> \> \mu{\rm eV} \> .
\label{final mu shift}
\eea
The first value is one order of magnitude below the present experimental 
accuracy  ($840$ Hz) in the $2S - 1S$ transition \cite{Exp}
whereas the planned Lamb shift experiment in muonic hydrogen 
\cite{PSI-R-98-03} aims for
a precision which is comparable to the uncertainty in $ \Delta E_{2S}^{\mu}$.
Incidentally, the proton polarization contribution to this Lamb shift has
nearly the same magnitude as the hadronic vacuum-polarization correction
\cite{KiNi} which, however, is more precisely known. A more accurate 
evaluation of the former contribution is therefore needed for a better
determination of the proton radius from the muonic Lamb shift experiment.

\vspace{0.5cm} 
\noindent
{\bf Note added:} After submission of the manuscript
additional calculations of the muonic polarization shift using different methods 
have been reported. Faustov and Martynenko \cite{FaMa} obtain nearly the
same value as reported in this paper whereas Pachucki's number \cite{Pach}
is slightly lower.

\vspace{0.5cm}
\noindent
{\bf Acknowledgements :} I would like to thank David Taqqu for inspiring
questions and many discussions. Thanks to Valeri Markushin for a
critical reading of the manuscript. Useful correspondence with I. B. Khriplovich
and R. A. Sen'kov about their calculation is acknowledged. Finally 
I am indebted to Simon Eidelman for additional information and to
Krzysztof Pachucki for sending me a draft of his manuscript before publication.


\vspace{1cm}


\begin{thebibliography}{99}

\bibitem{Exp} T. Udem {\it et al.}, Phys. Rev. Lett. {\bf 79} (1997) 2646;
B. de Beauvoir {\it et al.}, Phys. Rev. Lett. {\bf 78} (1997) 440; 
A. Huber {\it et al.}, Phys. Rev. Lett. {\bf 80} (1998) 468

\bibitem{deutpol} J. L. Friar and G. L. Payne, Phys. Rev. {\bf C 56} (1997) 
619 ; W. Leidemann and R.~Rosenfelder, Phys. Rev. {\bf C 51} (1995) 427;
J. Martorell, D. W. Sprung and D. C. Zheng, Phys. Rev. {\bf C 51} (1995) 1127;
Y. Lu and R. Rosenfelder, Phys. Lett. {\bf B 319} (1993) 7; 
K. Pachucki, D. Leibfried and T. W. H\"ansch, Phys. Rev. {\bf A 48}
(1993)  R1

\bibitem{KhSe} I. B. Khriplovich and R. A. Sen'kov, nucl-th/9704043, 
Phys. Lett. {\bf A 249} (1998) 474

\bibitem{PSI-R-98-03} F. Kottmann {\it et al.}, PSI-proposal R-98-03 
(December 1998)

\bibitem{BeEr} J. Bernab\'eu and T. E. O. Ericson, Z. Phys. {\bf A 309} (1983)
213

\bibitem{Ber} J. Bernab\'eu, private communication

\bibitem{Mainz} M. MacCormick {\it et al.}, Phys. Rev. {\bf C 53} (1996) 41 

\bibitem{Daresbury} T. A. Armstrong {\it et al.}, Phys. Rev. {\bf D 5} (1972) 
 1640 

\bibitem{Han} O. Hanstein {\it et al.}, Nucl. Phys. {\bf A 632} (1998) 561

\bibitem{BGM} D. Babusci, G. Giardano and G. Matone, Phys. Rev. 
{\bf C 57} (1998) 291 

\bibitem{DaGi} M. Damashek and F. J. Gilman, Phys. Rev. {\bf D 1} (1970) 1319

\bibitem{FMS} R. N. Faustov, A. P. Martynenko and V. A. Saleev, hep-ph/9811514

\bibitem{Simon} G. G. Simon {\it et al.}, Nucl. Phys. {\bf A 333} (1980) 381

\bibitem{Ros} R. Rosenfelder, Nucl. Phys. {\bf A 393} (1983) 301 

\bibitem{Holstein} B. Holstein, hep-ph/9710548

\bibitem{BHY} A. Buchmann, E. Hern\'andez and K. Yazaki, Nucl. Phys. 
{\bf A 569} (1994) 661

\bibitem{KiNi} T. Kinoshita and M. Nio, Phys. Rev. Lett. {\bf 82} (1999) 3240

\bibitem{FaMa} R. N. Faustov and A. P. Martynenko, hep-ph/9904362

\bibitem{Pach} K. Pachucki, private communication

\end{thebibliography}
\end{document}